\documentclass[12pt]{iopart}  
\pdfoutput=1

\usepackage{iopams}
\usepackage{graphicx}
\usepackage{subfigure}

\begin{document}
\title[Neutron capture on $^{130}$Sn during $r$-process freeze-out]{Neutron capture on $^{130}$Sn during $r$-process freeze-out}
\author {J. Beun$^1$, J. C. Blackmon$^2$, W. R. Hix$^3$, G. C. McLaughlin$^1$, M. S. Smith$^3$, R. Surman$^4$}
\address{$^1$ Department of Physics, North Carolina State University, Raleigh, NC 27695-8202}
\address{$^2$ Department of Physics, Louisiana State University, Baton Rouge, LA 70803}
\address{$^3$ Physics Division, Oak Ridge National Laboratory, Oak Ridge, TN 37831-6374}
\address{$^4$ Department of Physics, Union College, Schenectady, NY 12308}
\ead{jbbeun@ncsu.edu}

\begin{abstract}
We examine the role of neutron capture on $^{130}$Sn  during 
$r$-process freeze-out in the neutrino-driven wind environment of the core-collapse supernova.
We find that the global $r$-process abundance pattern
is sensitive to the magnitude of the neutron capture cross section of $^{130}$Sn.
The changes to the abundance pattern include 
not only a relative decrease in the abundance of $^{130}$Sn and an increase in the abundance of $^{131}$Sn,
but also a shift in the distribution of material in the rare earth and third peak regions.

\end{abstract}

\maketitle

\section{Introduction}
\label{rate_background}

Experiments at the Holifield Radioactive Ion Beam Facility (HRIBF) at Oak Ridge National Laboratory (ORNL)
are currently underway to acquire the nuclear data necessary to help generate precision neutron
capture cross sections for neutron-rich nuclei near the $N=82$ closed shell.
This data is relevant to the astrophysical community,
since the nucleosynthetic origin of most of the nuclei heavier than the Fe group ($Z\gtrsim26$)
arises from processes involving neutron capture.  
In particular, the neutron capture rates for the $N=82$ closed shell operate during the rapid
neutron capture process, known as the $r$-process \cite{1957RvMP...29..547B,1957Cameron}.

The classical $r$-process proceeds primarily through two phases, 
first, an equilibrium phase marked by $(n,\gamma) \rightleftharpoons(\gamma,n)$ equilibrium,
and then a freeze-out phase marked by the interplay between neutron capture
and photo-dissociation.
Early in the $r$-process, both neutron capture and photo-disintegration are fast,
hence $(n,\gamma) \rightleftharpoons(\gamma,n)$ equilibrium is established,
and the abundance of the nuclear species along each isotopic chain can be
found through the Saha equation,
for a review see Cowan et al. 1991 \cite{1991PhR...208..267C}.
At late times in the $r$-process, $(n,\gamma) \rightleftharpoons(\gamma,n)$ equilibrium fails
as the abundance of free neutrons is depleted
and the neutron-to-seed ratio of the $r$-process falls below one, $R\lesssim 1$.
During this freeze-out epoch of the $r$-process, 
individual neutron capture and photo-disintegration reactions now 
play a role in determining the final $r$-process abundance pattern.

\begin{figure}[h!]
\centering 
\includegraphics[width=\textwidth]{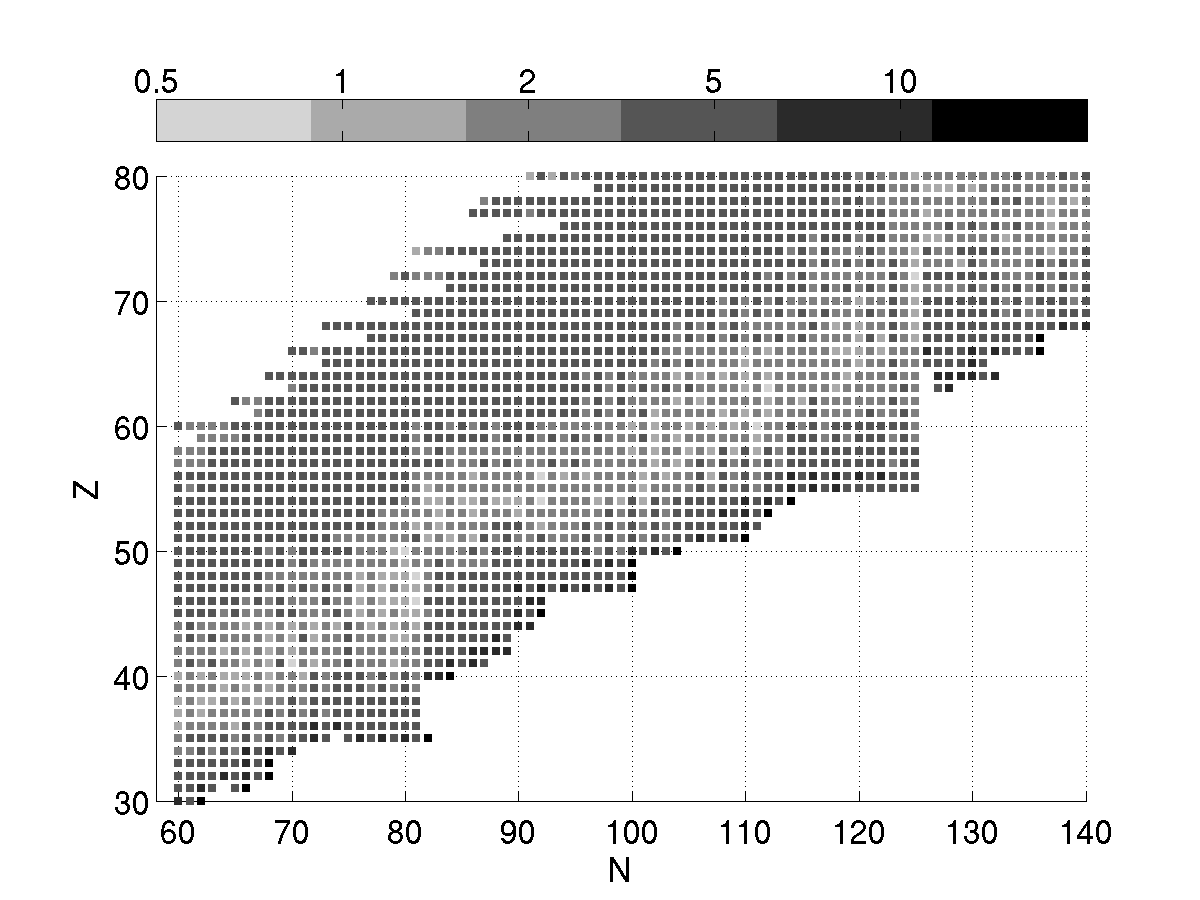}
\caption{We show the ratio (logarithmic) of the largest neutron capture cross section
with respect to the smallest neutron capture cross section 
between three sets of theoretical neutron capture cross sections.
The neutron capture cross sections can vary by many orders of magnitude 
between the theoretical models
(ratios $\gtrsim 10^{10}$ for the darkest squares).
}
\label{rcap_compare} 
\end{figure}

The relevant nuclear data for many of the large number of nuclei ($\gtrsim 3000$)
which participate in the $r$-process is sparse,
see {\emph e.g.} Pearson et al. 2006 \cite{2006NuPhA.777..623P}.
This is particularly true for the neutron capture cross sections.
Calculations of the $r$-process employ primarily phenomenological models of neutron capture cross sections
that use theoretical nuclear mass models as inputs.
The neutron capture cross section of a single nucleus can vary by
several orders of magnitude between individual theoretical models,
and an example of this is shown in \fref{rcap_compare} 
for three sets of theoretical neutron capture cross sections
based on the SMOKER model \cite{1991PhR...208..267C}, the FRDM model \cite{1995ADNDT..59..185M},
and the ETFSI model \cite{1995ADNDT..61..127A}.
To date there has been little calibration with experiment.

Proposed environments for the $r$-process include the neutrino-driven wind of a core-collapse supernova 
\cite{1992ApJ...395..202W,1992ApJ...399..656M,1994ApJ...433..229W,1994A&A...286..857T,1996ApJ...471..331Q,
1997ApJ...486L.111C,2000ApJ...533..424O,2000PASJ...52..601S,2001ApJ...554..578W,2001ApJ...562..887T,2008arXiv0805.1848P},
a prompt explosion from a low mass supernova \cite{2001ApJ...562..880S,2003ApJ...593..968W},
compact object mergers \cite{1999ApJ...525L.121F,2005NuPhA.758..587G,Surman:2008qf},
a gamma ray burst accretion disk \cite{2004ApJ...603..611S,Surman:2005kf,2005NuPhA.758..189M},
a collapsar from a massive stellar progenitor \cite{1999ApJ...524..262M,2004ApJ...606.1006P},
and the shocked surface layers of the post-collapse O-Ne-Mg Cores \cite{Ning:2007tu,Janka:2007di}.
No single site containing all of the essential ingredients for the $r$-process
has been fully realized, however. 
As such, many recent studies of the $r$-process have focused on 
understanding how an $r$-process would obtain 
given the current understanding of the astrophysical conditions.
Although the freeze-out epoch of the $r$-process presents a smaller contribution to the final $r$-process abundance pattern 
than that of the $(n,\gamma) \rightleftharpoons(\gamma,n)$ equilibrium epoch,
knowledge of the details of the dynamics of $r$-process freeze-out 
will be necessary for understanding the fine structure of the abundance pattern
as a clearer picture of the astrophysical environment of the $r$-process begins to emerge.

Previous works that have examined the role of neutron capture cross sections in the $r$-process
follow two main strategies.
Surman and Engel 2001 \cite{PhysRevC.64.035801} showed
that individual neutron capture cross sections in the region of $A \sim 195$ can influence the $r$-process
in this same region during freeze-out. 
Other works have examined global changes to the cross sections
\cite{rauscher-2005-758,2006AIPC..819..419F,1998PhLB..436...10G,1997A&A...325..414G,2004rpao.conf...63R}, 
and, in particular, Rauscher 2005 \cite{rauscher-2005-758}
examined how simultaneously changing all the $r$-process neutron capture cross sections
can influence the time at which $r$-process freeze-out begins.

We explore how an increase to the neutron capture cross section of $^{130}$Sn
can yield global changes to the abundance pattern of the $r$-process
compared to an $r$-process calculation under 
the original, unaltered neutron capture cross sections.
Experiments at HRIBF are in progress to collect data for $^{130}$Sn for the first time \cite{2007APS..DNP.HD010K}.
We organize this paper as follows.
In \sref{rate_sec_model}, we describe the details of our $r$-process calculation in the neutrino-driven wind.
\Sref{sec-second} describes the factors that enable $^{130}$Sn to
become influential to the $r$-process and the physics of the changes to the abundance pattern incurred by increasing the
$^{130}$Sn neutron capture cross section.
{{Additionally,
in this section we contrast our results for $^{130}$Sn with another nucleus under experimental examination, $^{132}$Sn
\cite{2007APS..APR.R2002J}.}}
We conclude in \sref{sec-fourth}.

\section{Model of $r$-process Nucleosynthesis}
\label{rate_sec_model}

Using the conditions of the neutrino-driven wind of the core-collapse
supernova environment as our guide,
we track the evolution of a mass element from the surface of the protoneutron star
in order to examine the effects of individual neutron capture cross sections at late times in the $r$-process.
We employ a 1D thermodynamic model of the neutrino driven wind as described in \cite{2006PhRvD..73i3007B,Beun:2007wf}.
We choose a set of wind conditions that yield the production of $r$-process elements through the third peak.
Unless stated otherwise, our calculations employ an entropy per baryon of $s/k = 100$, 
an outflow timescale of $\tau = 0.3 \, \ {\rm s}$,
and an initial electron fraction of $Y_e = 0.26$ at $T_9 = 10$.

\begin{figure}[h!]
\centering 
\includegraphics[trim = 0.0in 0.0in 0.0in 0.2in, clip, width=0.6\textwidth]{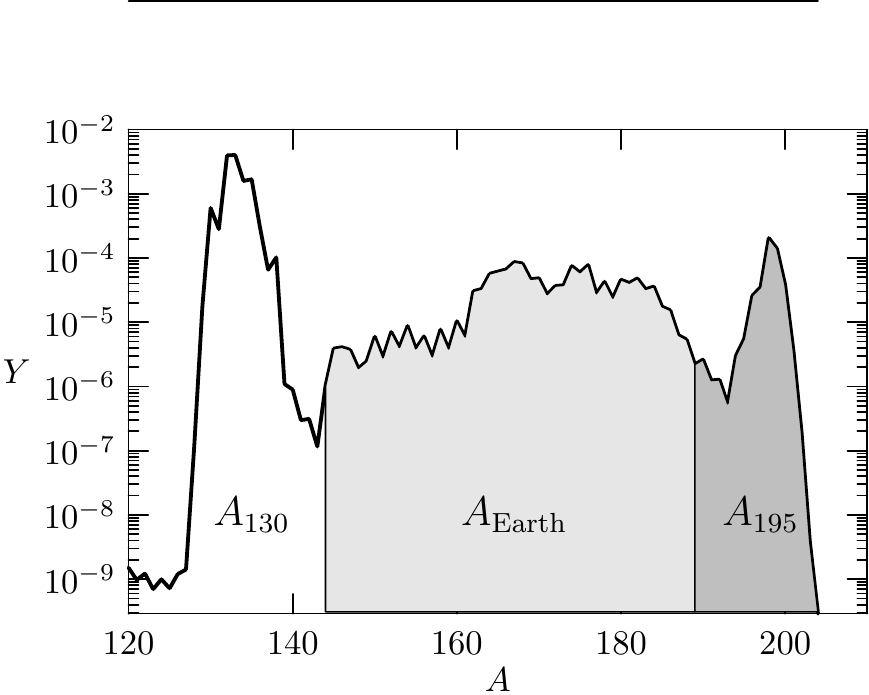}
\caption{We show the $r$-process abundance pattern in the neutrino-driven wind
for an entropy per baryon of $s/k = 100$, 
an outflow timescale of $\tau = 0.3 \, \ {\rm s}$,
and an initial electron fraction of $Y_e = 0.26$ at $T_9 = 10$.
The shaded areas depict the various peak regions of the $r$-process.
The left region (white) shows the second $r$-process peak,
the middle region (light gray) shows the rare earth peak region,
and the right region (dark gray) shows the third $r$-process peak region.}
\label{fig:abun_region} 
\end{figure}

To model the nucleosynthesis outcome, 
we employ two coupled reaction networks to track the abundance of the ejected mass element.
While the mass element is near the surface of the proto-neutron star, $T_9 \lesssim 10$, 
we employ an intermediate network calculation \cite{hix1999} 
that includes the strong and electromagnetic rates found in \cite{2000ADNDT..75....1R}. 
Once material reaches the $r$-process epoch, $T_9 \approx 2.5$,
we transition to an $r$-process network \cite{2001PhRvC..64c5801S,PhysRevLett.79.1809}
that includes the relevant reactions of $\beta$-decay, $\beta$-delayed neutron emission, neutron capture, and photo-disintegration.
We use $\beta$-decay rates from \cite{1990NuPhA.514....1M}, neutron separations from \cite{2003PhRvC..67e5802M},
and neutron capture rates from \cite{2000ADNDT..75....1R}.
Additional details of the network calculation can be found in \cite{2006PhRvD..73i3007B}.
In this initial study, as the changes to the $r$-process pattern occur late in the 
nucleosynthesis epoch, we do not include neutrino interactions
other than to set the initial electron fraction.
The $r$-process abundance pattern produced under these conditions in the neutrino-driven wind is shown in \fref{fig:abun_region}.

\section{$^{130}$Sn (n,$\gamma$) at Late Times}
\label{sec-second}

\begin{figure}[h!]
\centering 
\includegraphics[width=0.4\textwidth]{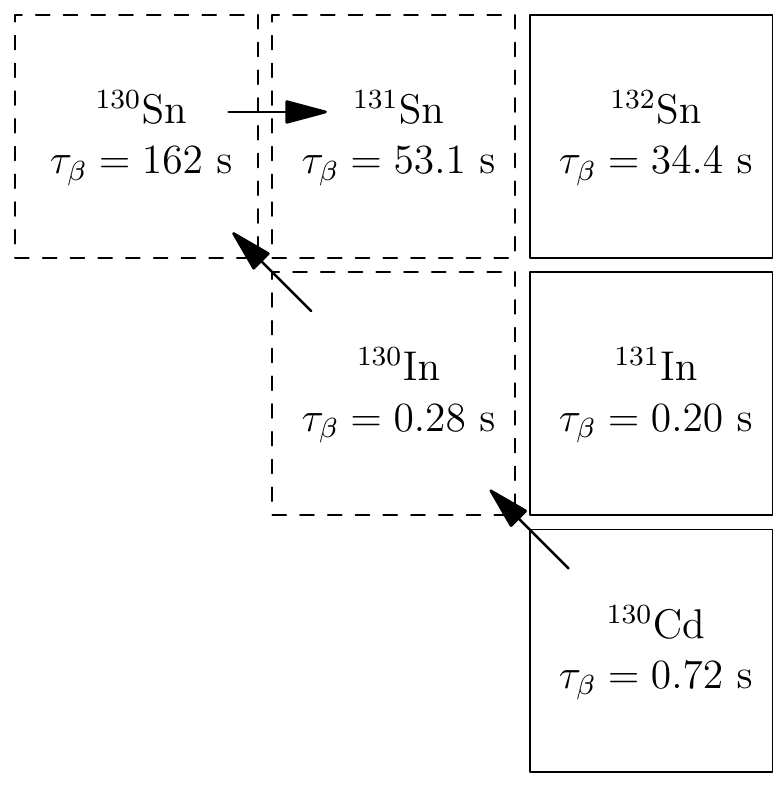}
\caption{ We depict the flow of material near $^{130}$Sn.
Rightward arrows indicate the direction
of the neutron capture reaction channel, and
the arrows that point along the diagonal indicate the direction 
of $\beta$-decay.
A significant amount of material arrives at $^{130}$Sn from the $\beta$-decay of the $A=130$ path nucleus $^{130}$Cd.
Depending on the neutron capture cross section of $^{130}$Sn,
some fraction of this material is re-directed along the $Z=50$
neutron capture reaction channel.
The conditions here are the same as in \fref{fig:abun_region}.}
\label{path_n_plot} 
\end{figure}

Near the end of $r$-process freeze-out,
while the free neutron abundance is rapidly falling, 
but neutron capture continues to be a relevant reaction in the $r$-process,
$^{130}$Sn becomes highly populated as material flows along the $A=130$ $\beta$-decay reaction pathway.
This pathway is significant as $^{130}$Cd is a closed shell nucleus
which is highly populated and has a short $\beta$-decay lifetime,
along with its daughter nucleus $^{130}$In.
The fraction of material that the now highly populated $^{130}$Sn 
directs along the $Z=50$ nuclei by neutron capture
is sensitive to the neutron capture cross section of $^{130}$Sn.
This sensitivity arises as a consequence of the long $\beta$-decay lifetime of $^{130}$Sn, $\tau_{\beta} = 162 {\rm  {\ s}}$,
which allows only relatively small amounts of material to exit $^{130}$Sn by $\beta$-decay.
These factors cause significant amounts of material to be directed 
primarily along the neutron capture channel of $^{130}$Sn, and
we demonstrate both the flow of material into $^{130}$Sn by $\beta$-decay
and the flow of material out of $^{130}$Sn by neutron capture in \fref{path_n_plot}.

We now examine how the uncertainties in the neutron capture cross
section of $^{130}$Sn affect the neutron capture reaction 
channel of $^{130}$Sn and the consequences that these uncertainties present to the $r$-process.
{{Although the neutron capture cross section of $^{130}$Sn is fairly consistent
between the theoretical models that are depicted in \fref{rcap_compare},
our understanding of the cross section is not complete.
{{For example, Rauscher et al. 1997 examined the direct-capture portion 
of the capture cross section of neutrons on nuclei and found that the direct-capture cross section varies 
by several orders of magnitude between the different theoretical models for $^{130}$Sn \cite{1997NuPhA.621..327R}. }}
In order to probe the influence of the uncertainties in the cross section,
we introduce a scaling factor, $\chi$, which modifies the $^{130}$Sn neutron capture cross section
\begin{equation}\label{e:scaling}
{\langle v\sigma_{n,\gamma}({^{130}{\rm {Sn}}})\rangle}^{\prime} =
\chi  {\langle v\sigma_{n,\gamma}({^{130}{\rm {Sn}}})\rangle}
\end{equation}
Above, $\langle v\sigma_{n,\gamma}({^{130}{\rm {Sn}}})\rangle$ is the
original, thermally averaged, neutron capture cross section of $^{130}$Sn,
${\langle v\sigma_{n,\gamma}({^{130}{\rm {Sn}}})\rangle}^{\prime}$ is the
modified neutron capture cross section of $^{130}$Sn,
and $\chi$ is a scalar multiplication factor.
Changes to the neutron capture cross section of $^{130}$Sn 
induce a corresponding change in the photo-dissociation rate, $\lambda_\gamma$, of $^{131}$Sn,
\begin{equation}\label{e:gamma_scaling}
\lambda^{\prime}_\gamma(^{131}{\rm {Sn}}) = \chi  \lambda_\gamma(^{131}{\rm {Sn}})
\end{equation}
Above, $\lambda_{\gamma}(^{131}{\rm {Sn}}$) is the original photo-dissociation
rate of $^{131}$Sn
and $\lambda^{\prime}_{\gamma}(^{131}{\rm {Sn}}$) is the modified photo-dissociation rate of $^{131}$Sn.

We examine the $r$-process in the neutrino driven wind for 
increases to the neutron capture cross section by a factor of 10 and 100 ($\chi=10$, 100).
For these cases the photo-dissociation rate of $^{131}$Sn is also increased by a factor of 10 and 100 respectively.
We take the $r$-process that forms in the neutrino-driven wind 
under the conditions that are described in \sref{rate_sec_model} as our baseline for comparison. 

Under an increase in the neutron capture cross section of $^{130}$Sn,
changes in the $r$-process abundance pattern relative to the baseline
occur not only for nuclei near $^{130}$Sn,
but also throughout the $r$-process abundance pattern.
Changes to the abundance near $^{130}$Sn due to 
an increase in the neutron capture cross section of $^{130}$Sn by a factor of 100 
include a decrease in the abundance of the $A=130$ nuclei by $\sim 93\%$
and an increase in the abundance of the $A=131$ nuclei by $\sim 191\%$.
Global effects to the abundance pattern
include changes to the abundance of individual nuclear species
above the second $r$-process peak region by as much as $\sim 34\%$.
The average percent change in abundance for this increase in the cross section is $\sim 12\%$.
Although the percent change in abundance is less when
the neutron capture cross section of $^{130}$Sn is increased by a factor of 10,
global changes to the abundance pattern still occur.
The impact an increase in the neutron capture cross section of $^{130}$Sn has on 
the $r$-process abundance pattern is shown in \fref{ab_compare}.
\begin{figure}[h!]
\begin{center}
    \subfigure[]{\label{fig:ab_comp_10}\includegraphics[width=0.45\textwidth]{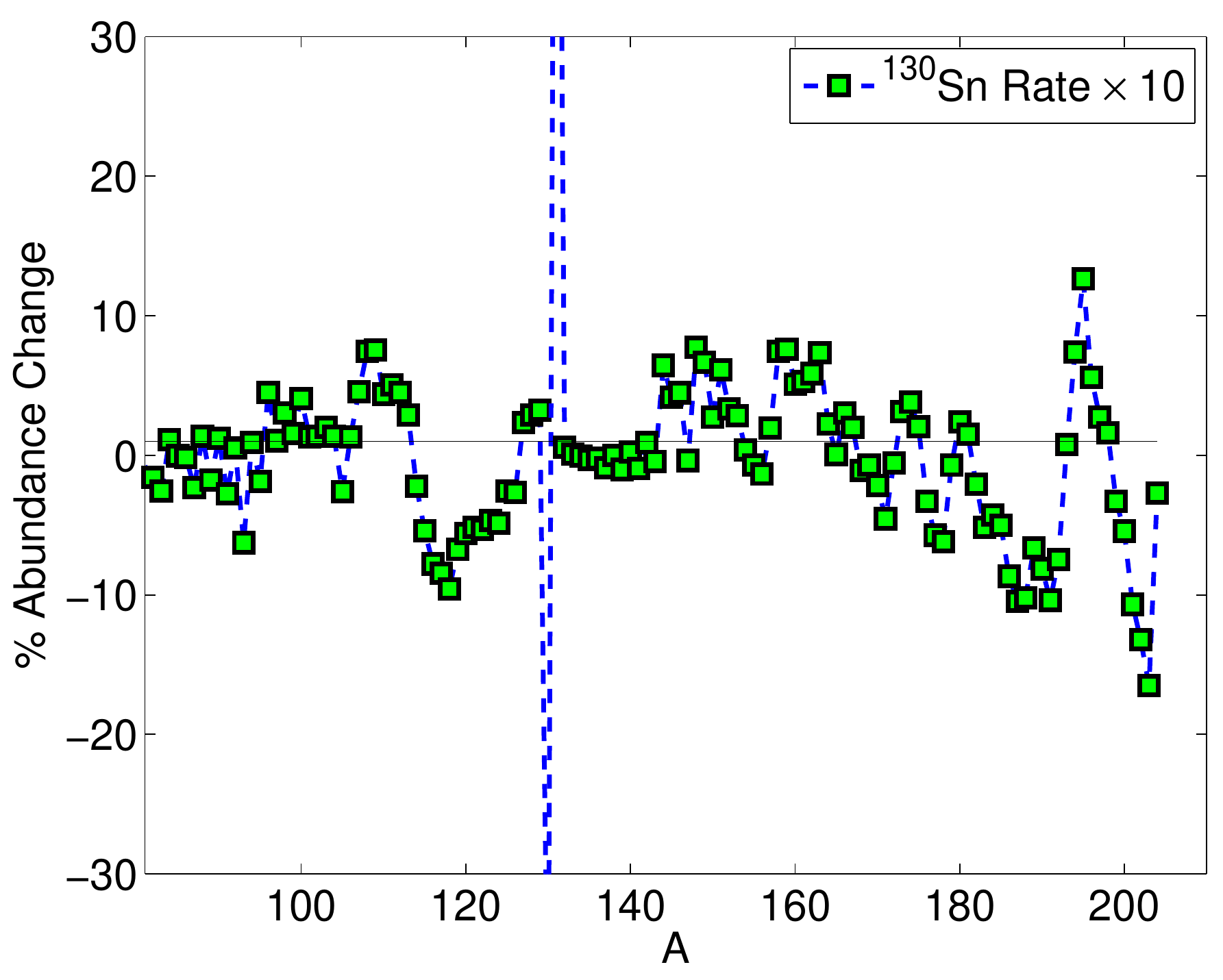}}
    \subfigure[]{\label{fig:ab_50_132__10}\includegraphics[width=0.45\textwidth]{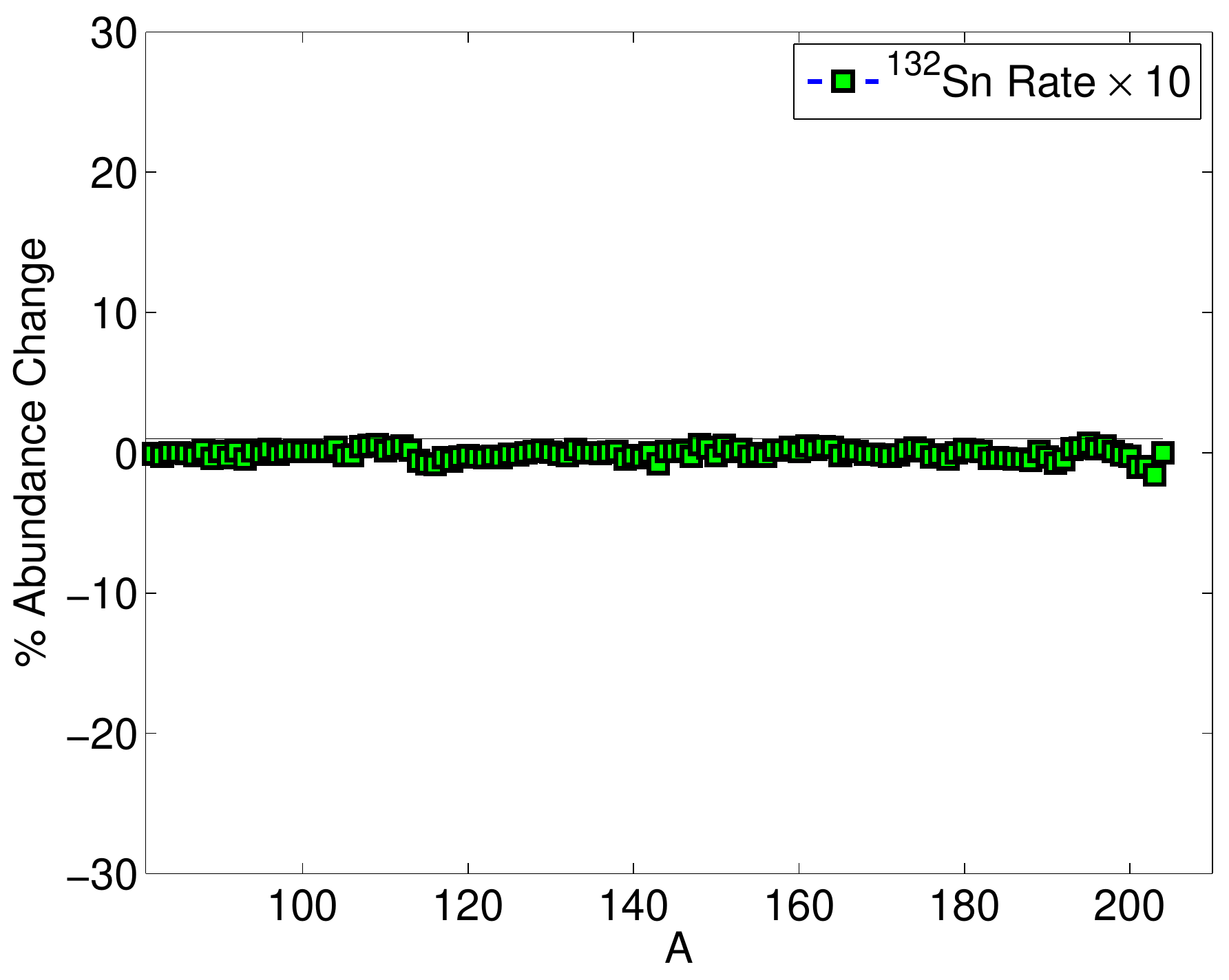}} \\
    \subfigure[]{\label{fig:ab_comp_}\includegraphics[width=0.45\textwidth]{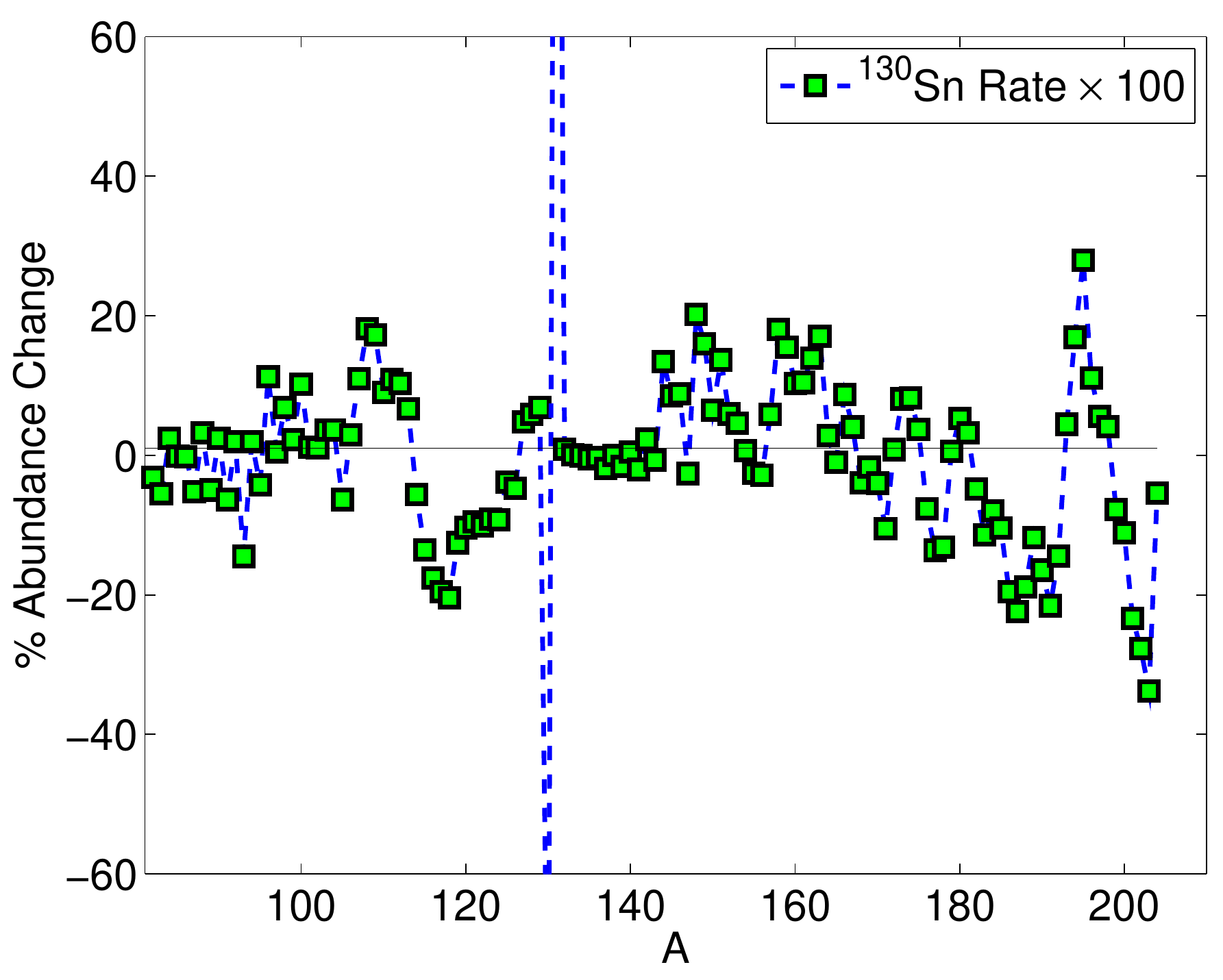}}
    \subfigure[]{\label{fig:ab_50_132__100}\includegraphics[width=0.45\textwidth]{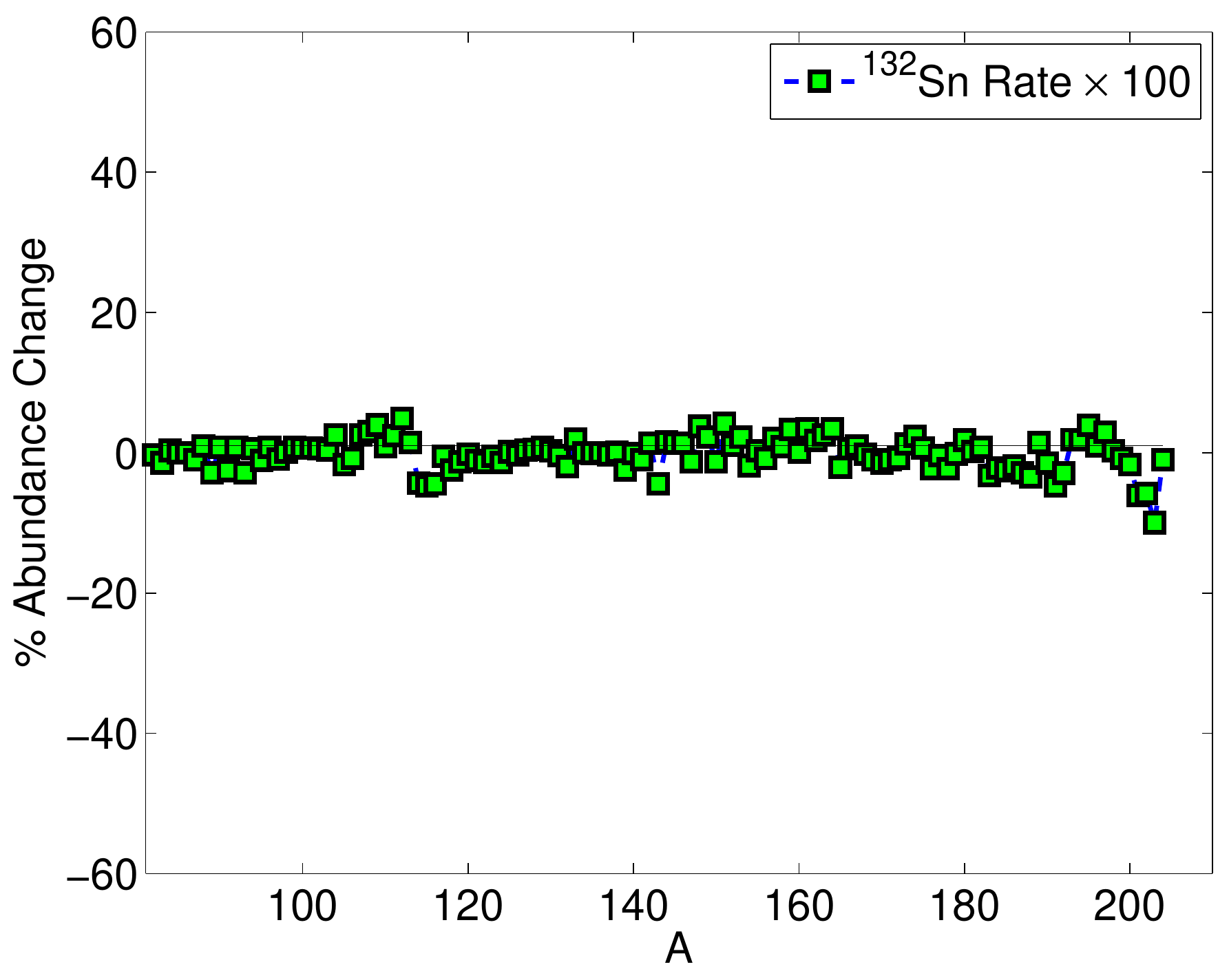}} 
\end{center}
\caption{The percent change in abundance for an $r$-process calculation under an increase in the
neutron capture cross section of $^{130}$Sn with respect to the baseline calculation is shown.
The top left panel depicts the percent change when the neutron capture cross section of $^{130}$Sn
is increased by a factor of 10, and the bottom left panel depicts the percent change for an increase by a factor of 100.
{{This is contrasted with the percent change in abundance due to changing 
the neutron capture cross section of $^{132}$Sn by a factor of 10, top right panel, and by a factor of 100, bottom right panel.}}
} 
\label{ab_compare} 
\end{figure}

\begin{figure}[h!]
\begin{center}
    \subfigure[]{\label{nout_130sn}\includegraphics[width=0.5\textwidth]{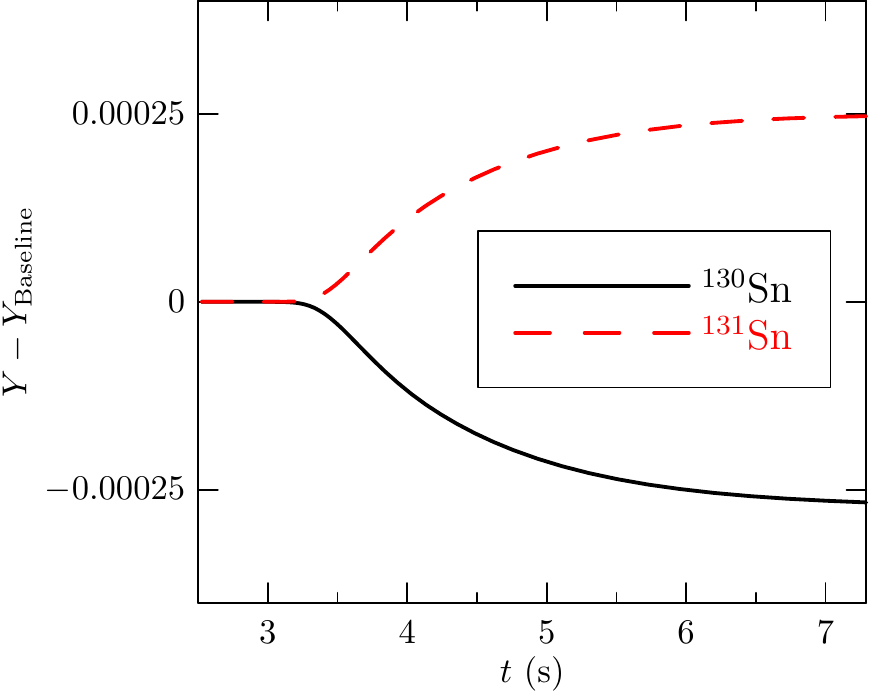}} \\
    \subfigure[]{\label{nout_global}\includegraphics[width=0.5\textwidth]{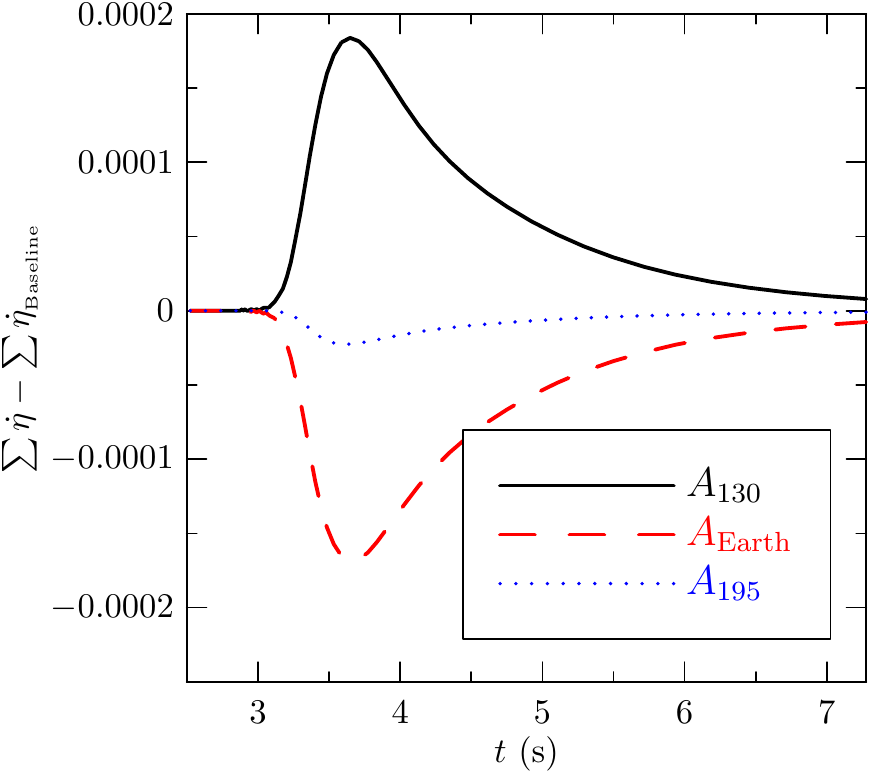}}
\end{center}
\caption{In the bottom panel,
we show the relative rate of net neutron capture, $\dot \eta$,
summed over each of the three peak regions of the $r$-process.
The absolute difference between an $r$-process with the neutron capture cross section of $^{130}$Sn increased by a factor of 100,
$\sum \dot \eta$, 
and that of the baseline calculation, $\sum \dot \eta_{_{{\rm {Baseline}}}}$, is shown.
The increase in the neutron capture cross section of $^{130}$Sn enhances
the amount of neutron capture that occurs in the $A \sim 130$ peak and 
reduces the amount of neutron capture for nuclei in the rare earth and $A \sim 195$ peak regions.
In the top panel, we show the absolute difference in the abundance of $^{130}$Sn (solid line) 
and $^{131}$Sn (dashed line)
for an $r$-process with the neutron capture cross section of
$^{130}$Sn increased by a factor of 100, $Y$,
and the baseline $r$-process calculation with 
the unaltered set of neutron capture rates, $Y_{\rm {Baseline}}$.
The depicted abundance change in the top panel is commensurate with that expected 
from increasing the $^{130}$Sn neutron capture rate.
} 
\label{fig:new_figs} 
\end{figure}

An increase of the $^{130}$Sn neutron capture cross section acts 
to hasten both the depletion of $^{130}$Sn and the production of $^{131}$Sn.
This is depicted in \fref{nout_130sn}.
A surprising consequence that results from increasing the cross section of $^{130}$Sn
is that enough neutrons are consumed in the production of $^{131}$Sn by the neutron capture on $^{130}$Sn
to significantly influence the availability of neutrons for capture by other $r$-process nuclei.
This effect can be quantified by the net rate of neutron capture,
\begin{eqnarray}
\dot{\eta}(Z,A) = n_n\langle v\sigma_{n,\gamma}(Z,A)\rangle Y(Z,A) - \lambda_{\gamma}(Z,A+1) Y(Z,A+1)
\label{net_capture_rate}
\end{eqnarray}
The net rate of neutron capture is the instantaneous rate of neutrons that are captured or released 
in $(n,\gamma)$ reactions between the nuclei $(Z,A)$ and $(Z,A+1)$.
The increase in the net rate of neutron capture for the second peak region
is consequentially balanced by an equivalent decrease in the net rate of neutron capture for the rare earth
and third peak regions.
This is shown in \fref{nout_global},
where we sum the net rate of neutron capture
over each of the $r$-process peak regions, $\sum \dot{\eta}$.
The individual $r$-process peak regions are the same as those depicted in \fref{fig:abun_region};
the $A=130$ peak region is taken to include nuclei with $A<145$,
the rare earth peak region includes nuclei between $145 \leq A < 190$,
and the $A=195$ peak region includes nuclei with $A \geq 190$

This reduction in the overall rate of neutron capture above the $A \approx 130$ $r$-process peak 
influences the distribution of material within the rare earth and third peak regions by shifting material within these
regions towards lighter nuclei in $A$.
For an $r$-process under the increased neutron capture cross section,
less material reaches the right hand side of the third $r$-process peak, in nuclei with $A \geq 199$,
compared with the abundance pattern of the baseline calculation.
Additionally, an increase in the abundance of the nuclei
is found on the left hand side of the third $r$-process peak in the nuclei between $195 \leq A \leq 198$.
A similar effect occurs in the rare earth region of the $r$-process,
less material populates the heavier, right hand side of the rare earth region,
leading to an increase in the abundance of lighter nuclei within the rare earth region.
The impact of the shifting of material in the $r$-process under an increased neutron capture cross section
of $^{130}$Sn is depicted in \fref{ab_compare}.

\subsection{Comparison with $^{132}$Sn (n,$\gamma$)}

{{ We contrast our findings for $^{130}$Sn
with a classical waiting point nucleus, $^{132}$Sn,
which is also being studied at HRIBF \cite{2007APS..APR.R2002J}.
This nucleus, $^{132}$Sn,
is both significantly populated at late times in the $r$-process and has a relatively long $\beta$-decay lifetime,
similar to $^{130}$Sn.
The dynamics of the material flows near $^{132}$Sn 
depart from that of $^{130}$Sn in a significant way, however.
The $(n,\gamma) \rightleftharpoons(\gamma,n)$ counterpart nucleus of $^{132}$Sn,
$^{133}$Sn, has a faster photo-dissociation rate than that of $^{131}$Sn, 
the $(n,\gamma) \rightleftharpoons(\gamma,n)$ counterpart to $^{130}$Sn.
This occurs despite the lower neutron capture rate of $^{132}$Sn relative to $^{130}$Sn
and is largely due to the smaller neutron separation energy of $^{133}$Sn 
($S_n \approx 2.6 {\rm {\ MeV}}$)
relative to that of $^{131}$Sn
($S_n \approx 5.1 {\rm {\ MeV}}$).
This increased rate of photo-dissociation results in $^{132}$Sn 
remaining in $(n,\gamma) \rightleftharpoons(\gamma,n)$ equilibrium 
until very late times in the $r$-process, 
which in this model corresponds to $t \approx 4.5 {\rm {\ s}}$.
After $^{132}$Sn leaves $(n,\gamma) \rightleftharpoons(\gamma,n)$ equilibrium,
the $r$-process does gain slight sensitivity to its neutron capture cross section, 
see \fref{ab_compare}.
The free neutron abundance is low as this happens late in the $r$-process,
leading to a smaller overall effect to the $r$-process than that for the case of $^{130}$Sn. }}

\section{Conclusions}
\label{sec-fourth}

We have shown how the neutron capture cross section of a single nucleus,
$^{130}$Sn, can influence the shape the global $r$-process abundance pattern.
Changes to the neutron capture cross section of $^{130}$Sn can influence the abundance pattern
at late times,
shifting material towards lighter nuclei in the rare earth and third peak regions 
for increases in the neutron capture cross section of $^{130}$Sn.
This arises as an increase in the cross section enhances the amount of
neutron capture by $^{130}$Sn and reduces the amount of neutron capture in the rare earth and third peak regions. 
For an increase in the neutron capture cross section of $^{130}$Sn by a factor of 100, 
we find changes to the abundance of nuclei near $^{130}$Sn are as much as $\sim 191\%$,
and changes to the abundance of nuclei near the third peak region are as much as $\sim 34\%$.
We find the average percent change in the abundance of the $r$-process nuclei to be $\sim 12\%$.
Although smaller increases to the cross section reduce the size of these effects,
we show that an increase in the neutron capture cross section of $^{130}$Sn by a factor of 10
results in similar global changes to the $r$-process abundance pattern.
These changes are specific to the astrophysical conditions of the $r$-process
and depend on the underlying nuclear data and choice of the thermodynamic model.
While the $(n,\gamma) \rightleftharpoons(\gamma,n)$ equilibrium epoch of the $r$-process
sets the major features of the abundance pattern of the $r$-process,
the details of the physics of the $r$-process freeze-out epoch,
including the neutron capture cross section of key nuclei such as $^{130}$Sn,
impacts the abundance pattern as well.
{{Experimental investigations of the neutron capture cross section of $^{130}$Sn
will help to reduce the uncertainties in this cross section,
and the initial efforts are underway at the Hollifield Radioactive Ion Beam Facility (HRIBF).}}

The influence of the neutron capture reaction for $^{130}$Sn demonstrated  
in this analysis results from the confluence of three factors; a  
prominent place on the $r$-process path, a long $\beta$-decay half-life,
and a large $Q$ value for the neutron capture.  The first two result in  
a significant abundance of $^{130}$Sn at late times in the $r$-process,
while the last causes this reaction to drop out of $(n,\gamma) \rightleftharpoons(\gamma,n)$ equilibrium 
early enough that significant numbers of neutron captures  
can occur out of equilibrium.  While the half-life and $Q$-value are  
universal, the dependence on the $r$-process path means that the  
influence demonstrated here for $^{130}$Sn may not hold for all $r$-process  
scenarios.  However, it is likely that other reactions, with similar  
characteristics, would have similar influence for these differing  
scenarios. Determination of the existence and identity of these  
influential reactions, as well as theoretical and experimental study  
of their rates, is important to refining our knowledge of the $r$-process.

\ack
This work was partially supported by the Joint Institute for Heavy Ion Research at ORNL,
the Department of Energy under contracts DE-FG05-05ER41398 (RS), DE FG02-02ER41216 (GCM),
 and by the National Science Foundation under contract PHY-0244783 (WRH).
Oak Ridge National Laboratory is managed by UT-Battelle, LLC, for the U.S. Department of Energy under 
contract DE-AC05-000R22725.


%
%
%
%


\makeatletter
\let\jnl@style=\rm
\def\ref@jnl#1{{\jnl@style#1}}
\def\ref@jnl#1{{#1}}
\def\ref@jnl{{\jnl@style}}

\def\aj{\ref@jnl{AJ}}                   
\def\araa{\ref@jnl{ARA\&A}}             
\def\apj{\ref@jnl{ApJ}}                 
\def\apjl{\ref@jnl{ApJ}}                
\def\apjs{\ref@jnl{ApJS}}               
\def\ao{\ref@jnl{Appl.~Opt.}}           
\def\apss{\ref@jnl{Ap\&SS}}             
\def\aap{\ref@jnl{A\&A}}                
\def\aapr{\ref@jnl{A\&A~Rev.}}          
\def\aaps{\ref@jnl{A\&AS}}              
\def\azh{\ref@jnl{AZh}}                 
\def\baas{\ref@jnl{BAAS}}               
\def\jrasc{\ref@jnl{JRASC}}             
\def\memras{\ref@jnl{MmRAS}}            
\def\mnras{\ref@jnl{MNRAS}}             
\def\pra{\ref@jnl{Phys.~Rev.~A}}        
\def\prb{\ref@jnl{Phys.~Rev.~B}}        
\def\prc{\ref@jnl{Phys.~Rev.~C}}        
\def\prd{\ref@jnl{Phys.~Rev.~D}}        
\def\pre{\ref@jnl{Phys.~Rev.~E}}        
\def\prl{\ref@jnl{Phys.~Rev.~Lett.}}    
\def\pasp{\ref@jnl{PASP}}               
\def\pasj{\ref@jnl{PASJ}}               
\def\qjras{\ref@jnl{QJRAS}}             
\def\skytel{\ref@jnl{S\&T}}             
\def\solphys{\ref@jnl{Sol.~Phys.}}      
\def\sovast{\ref@jnl{Soviet~Ast.}}      
\def\ssr{\ref@jnl{Space~Sci.~Rev.}}     
\def\zap{\ref@jnl{ZAp}}                 
\def\nat{\ref@jnl{Nature}}              
\def\iaucirc{\ref@jnl{IAU~Circ.}}       
\def\aplett{\ref@jnl{Astrophys.~Lett.}} 
\def\apspr{\ref@jnl{Astrophys.~Space~Phys.~Res.}}
\def\bain{\ref@jnl{Bull.~Astron.~Inst.~Netherlands}} 
\def\fcp{\ref@jnl{Fund.~Cosmic~Phys.}}  
\def\gca{\ref@jnl{Geochim.~Cosmochim.~Acta}}   
\def\grl{\ref@jnl{Geophys.~Res.~Lett.}} 
\def\jcp{\ref@jnl{J.~Chem.~Phys.}}      
\def\jgr{\ref@jnl{J.~Geophys.~Res.}}    
\def\jqsrt{\ref@jnl{J.~Quant.~Spec.~Radiat.~Transf.}}
\def\memsai{\ref@jnl{Mem.~Soc.~Astron.~Italiana}}
\def\nphysa{\ref@jnl{Nucl.~Phys.~A}}   
\def\physrep{\ref@jnl{Phys.~Rep.}}   
\def\physscr{\ref@jnl{Phys.~Scr}}   
\def\planss{\ref@jnl{Planet.~Space~Sci.}}   
\def\procspie{\ref@jnl{Proc.~SPIE}}   

\let\astap=\aap
\let\apjlett=\apjl
\let\apjsupp=\apjs
\let\applopt=\ao

\section{References}
\bibliographystyle{unsrt}
\bibliography{master}

\end{document}